# Controlled dephasing of an electron interferometer with a path-detector at equilibrium

E. Weisz , H. K. Choi , M. Heiblum , Yuval Gefen , V. Umansky and D. Mahalu

*Braun Center for Submicron Research, Department of Condensed Matter Physics,*

*Weizmann Institute of Science, Rehovot* 76100, *Israel*

## Abstract

Controlled dephasing of electrons, via 'which path' detection, involves, in general, coupling a coherent system to a current driven noise source. However, here, we present a case in which a nearly isolated electron puddle at thermal equilibrium strongly affects the coherence of a nearby electronic interferometer. Moreover, for certain average electron occupations of the puddle, the interferometer exhibits complete dephasing. This robust phenomenon stems from the Friedel Sum Rule, which relates a system's occupation with its scattering phases. The interferometer opens a peeping window into physics of the isolated electron puddle, which cannot be accessed otherwise.

*Introduction*

Dephasing of quantum systems is highly interesting both from a practical point of view, since controlling and preventing dephasing are paramount in the manipulation of quantum states, and from a theoretical point of view, being a building block of quantum theory. Thus, in recent years experiments involving controlled dephasing (i.e. with a highly controlled, artificial, dephasing environment) were used to explore the robustness of coherent systems [1, 2, 3, 4, 5, 6, 7, 8] In this paper, we report of a novel case of controlled dephasing of a two-path electronic Mach-Zehnder interferometer (MZI) [9] strongly coupled to an electron puddle confined within a quantum dot (QD). Tuning the unbiased QD to resonance transmission, with no current flowing, induced, unexpectedly, a complete visibility loss in the adjacent MZI; accompanied by an abrupt π phase jump across the zero visibility point. This dephasing process, which was found to be remarkably independent of system parameters, was attributed to a manifestation of the Friedel Sum Rule, which connects the occupation of a system to its scattering phases [10, 11]. Since the observed visibility pattern images the resonance peaks of the electron puddle, we were able to determine the average occupation, charge fluctuations, and the lifetime of the electrons without running current through it. In addition, the roles of the finite temperature and back-action will be discussed.

*Experimental Procedures*

Employing chiral edge modes transport in the integer quantum Hall effect (IQHE) regime, we realized a strongly coupled system of an electronic MZI and a QD – as shown schematically in Fig. 1a. At filling factor two (*ff*=2, two spin resolved occupied Landau levels), the outer edge mode, with chemical potential $\mu_{outer}$, splits to two paths that recombine to form the MZI [9], while the inner edge mode is fully reflected by the input quantum point contact of the MZI (QPC1). The outer modes, either the one that forms the upper path of the MZI (arriving from the source at $\mu_{outer}$), or the 'cold' one returning from the grounded contact, both traverse freely through the QD. The inner edge mode, with chemical potential of its source $\mu_{inner}$, tunnels resonantly through the QD. In turn, the QD is tuned to the Coulomb blockade regime for the inner edge mode [12]. Being confined within the QD, the inner edge mode forms an electron puddle in the center of the QD, surrounded, almost symmetrically, by the freely passing outer edge modes. Coupling between the outer and inner edge modes (outside and inside the QD) is capacitive with inter-tunneling not observable. When degeneracy between $N$ and $N+1$ electrons in the puddle takes place, the number of electrons in the puddle is allowed to fluctuate; otherwise, their number remains fixed and the QD is said to be in the Coulomb blockade regime. The strong Coulombic coupling between the inner puddle and the outer mode ensures extreme sensitivity of the outer mode trajectories to the puddle's potential and its occupancy – both controlled by the plunger gate (QDP) voltage. At our quantizing magnetic field, even a minute change of some 750nm$^2$ in the enclosed area between the two interfering paths suffices to alter the Aharonov-Bohm (AB) phase by $2\pi$ - making the MZI an extremely sensitive potential detector.

The system was implemented in a two dimensional electron gas (2DEG), embedded in a GaAs-AlGaAs heterostructure, with areal density $n=2.7\times10^{11}$cm$^{-2}$ and low temperature mobility $\mu=2.5\times10^{6}$cm$^2$/V·s. The device was fabricated using wet etching and Au/Ge/Ni ohmic contacts [13] defined by optical photolithography, as well as PdAu/Au surface gates and Ti/Au air bridges defined by electron beam lithography. The properties of the MZI are set by two QPCs, QPC1 and QPC2, playing the role of the two beam splitters in an optical MZI (Figs. 1a & 1b). The outer edge mode emanated from a far left source ohmic contact, split at QPC1 and recombined at QPC2, with an accumulated phase difference between the two paths governed by the enclosed AB flux 9, $\varphi_{AB}=2\pi AB/\phi_0$, with $A=6\times6\mu m^2$ the enclosed area, $B=5.6$T the magnetic field, and $\phi_0=h/e$ the flux quantum. The inner edge mode arrived from a different source contact, fully reflected from QPC1, and directed towards the QD. The phase of the MZI is modulated by changing its area via charging the modulation gate (MG). The QD is defined by four surface gates: 'fork' gate (QDF), 'left' gate (QDL), 'right' gate (QDR), and a 'plunger' gate (QDP); confining an area of $0.2\times0.2\mu m^2$ with some $N\approx100$ enclosed electrons. The surface gates QDL, QDR and QDF are tuned to form two barriers for the inner edge mode while allowing the outer edge mode to traverse the QD freely. The voltage on the QDP controls the potential within the QD, and thus the occupancy of the confined inner puddle, via $\Delta u=\alpha\Delta V_{QDP}$, where $\alpha$ is the so called 'levering factor' [12]. Electron addition spectrum is set by two energy scales: the 'classical' charging energy $U_c=e^2/C\sim100\mu eV$, with $C$ the total puddle capacitance and the single particle level spacing $\Delta\sim20\mu eV$.

The sample was cooled in a dilution refrigerator with electron temperature ~45mK. The differential conductance was measured by applying a 3μV RMS at 865kHz excitation signal, with the drain signal filtered by a narrow band *LC* circuit; amplified by a two stage amplification chain with gain ~1000 (a cryogenic preamplifier at the 1.2K stage followed by a room temperature amplifier); measured by a spectrum analyzer connected to D1. Although each edge mode could be charged separately (by injecting from two sources and partially pinching QPC0 to fully transfer one mode and fully reflect the other one), they were monitored together at D1.

## *Results*

We start with the QD fully open and QPC1 and QPC2 of the MZI tuned each to transmission $t=1/2$ of the outer edge mode, while fully reflecting the inner mode. The signal, *I*, was monitored at D1 as function of the AB phase, controlled via $V_{MG}$. The observed visibility was $v\sim10\%$, with $v = \frac{I_{max} - I_{min}}{I_{max} + I_{min}}$.

We begin with a slightly pinched QD, with Coulomb blockade peaks in the conductance of the inner edge modes, while the outer modes pass freely (Fig. 2b). The conductance peaks gradually vanished as the coupling of the puddle to the leads diminished. Via biasing the impinging inner mode the non-linear conductance evolved into the familiar 'diamond-like' picture [12], leading to a levering factor α~0.0025 for the slightly pinched dot. As the QDP is negatively charged, it depleted the electrons puddle – as expected - however, due to inadvertent electrostatic coupling between the QDP and QDL & QDR, the QD gradually pinched and its resonant levels FWHM, Γ, narrowed. As Γ approached the temperature, the conductance peaks diminished as $\sim\Gamma/k_BT$. Now, the AB oscillations of the MZI were monitored as function of the QDP voltage. The bare interference pattern measured in D1 is presented by a color plot, as function of the MG and QDP voltages (Fig. 2a). Subtracting the inadvertent effect of the QDP on the area of the interferometer and compensating for the natural decay of the magnetic field during the time of the measurement (the superconducting coil, being in 'persistent mode', losses continuously quantum fluxes), the visibility and phase of the AB oscillation (extracted by a discreet fast Fourier transform) were plotted in Figs. 2c and 2d. In a periodic manner, reflecting the series of conductance peaks (Fig. 2b), the visibility exhibited a series of dips with the interference phase evolving monotonically throughout the Coulomb blockade valleys, lapsing by a nearly similar phase throughout the conductance peaks. As shown in the inset in Fig. 2b, the visibility dip (plotted in dots) mimics exactly the Lorenzian shaped conductance peaks (plotted in solid line). We wish to stress that the behavior of the visibility and phase remain the same when the inner edge mode was *not biased*, with the QD in thermal equilibrium.

We explored further the partial dephasing of the MZI by pinching the QD further, assuring nearly full reflection of the inner edge mode from the dot, while still fully passing the outer mode. Under these conditions, the visibility dropped periodically to nearly zero, with phase evolving by π between the visibility dips, lapsing abruptly by π at the visibility minima. In Fig. 3c an expanded view of a single visibility dip, and the associated phase, are shown. Note that this behavior was found to be robust and independent of magnetic field, the electrostatics of the QD and the device under study. In this range the levering factor was determined to be α=0.015.

Before we turn to discuss the implications of these results, we further explore the QD-MZI interaction in the non linear regime, yet under the conditions of Fig. 2 where the QD conducted the inner edge mode. Biasing the inner edge mode opened a window for $\mu_{QD}$, $\mu_{inner}>\mu_{QD}>0$, for which current traversed the puddle. For a left-right symmetrically tuned QD, we observed a visibility drop - as function of $V_{QDP}$ - throughout the bias window (Fig. 4a), accompanied by a phase change near $\mu_{inner}$ and 0 (Fig. 4b). By contrast, when the QD was tuned to a strong left-right asymmetry, the visibility drop and the phase change were observed only when $\mu_{QD}$ was aligned with the electrochemical potential of the better coupled lead – in this case, the source (Fig. 4c and 4d). We return to this point after delving into the mechanism of the dephasing process.

## Discussion

Addressing first the case of a pinched QD is more illuminating. The number of electrons in the puddle affects, via Coulomb interaction, the trajectory of electrons in the outer edge modes. Consequently, the induced AB phase $\varphi_{AB}(N)$ or $\varphi_{AB}(N+1)$ will depend on the puddle's occupancy. Since the dwell time of an electron in the confined puddle is much longer than the traversal time of an outer edge mode electron, each electron in the puddle interacts with many MZI electrons. We define $P_N$ ($P_{N+1}$) as the probability of having $N$ ($N+1$) electrons in the puddle, with $P_N+P_{N+1}=1$ and $P_N=P_{N+1}=0.5$ at the $N$-$N+1$ degeneracy point. Denoting $\Delta\varphi\equiv\varphi_{AB}(N+1)-\varphi_{AB}(N)$, the average output current of the MZI is $I \propto P_N\cos\varphi_{AB} + P_{N+1}\cos(\varphi_{AB}+\Delta\varphi)$ with visibility $v = \sqrt{P_N^2 + P_{N+1}^2 + 2P_N P_{N+1}\cos\Delta\varphi}$ and the actual MZI phase $\varphi_{MZI} = \tan^{-1}\left[\dfrac{P_{N+1}\sin\Delta\varphi}{P_N + P_{N+1}\cos\Delta\varphi}\right]$ [8]. Evidently, if $\Delta\varphi=\pi$ the AB oscillations will be fully quenched at the degeneracy point - as indeed was found for a rather pinched QD (see Fig 3c), when puddle's occupation changed from $N$ to $N+1$. Moreover, as the expression above suggests, the observed MZI phase, $\varphi_{MZI}$, underwent through an abrupt phase lapse of $\pi$ at the degeneracy point (see Fig. 3c). For $\Delta\varphi<\pi$, the visibility loss is partial and the phase evolution smoother at the degeneracy point – as was found for a slightly pinched QD (see Figs. 2c & 2d). This extremely sensitive detection of the dot's occupancy allowed a direct observation of the line shape of each resonance level, as reflected via $P_N$, even at thermal equilibrium and when the QD was fully pinched off. It is interesting to note that charging the QD negatively, say, via the applying a negative voltage to the QDP, affects the MZI phase in a somewhat counter-intuitive manner. Such charging will raise the energy of the Landau levels and thus move the intersections of the levels with the Fermi energy towards the center of the dot, hence *increasing* the AB flux in the MZI. An increased flux in the MZI induces a positive increase in the $\varphi_{AB}$ phase.

What is the justification in writing $\varphi_{AB}(N+1)-\varphi_{AB}(N)=\pi$ in a rather pinched dot, and why is it so robust? We propose that this is an unambiguous manifestation of the generalized 'Friedel Sum Rule' [10], which connects the scattering phases of a system with the number of occupied states. In a pinched QD, with unity transmission of the outer edge modes (the 'lower-hot' and 'upper-cold') and nearly full reflection of the inner edge modes (coming from the left and from the right; Fig 1a), the inner puddle is strongly capacitively coupled to the nearby outer edge modes and only weakly coupled to the inner, further away, inner modes. Hence, the scattering matrix $S$ of the QD involves only the two outer modes via $\psi_{outgoing}=S\psi_{incoming}$. The Friedel sum rule dictates $\Delta\varphi_{outer}(l\rightarrow r)+\Delta\varphi_{outer}(r\rightarrow l)=2\pi n$, where $n$ is the number of added electrons to the puddle [14]. For the puddle located symmetrically between the outer lower ($l\rightarrow r$) and the outer upper ($r\rightarrow l$) edge modes, the total scattering phase is equally shared between the two modes - with $\Delta\varphi_{MZI}=\pi$ for every added electron to the puddle. Hence, the robustness of the dephasing process emanates from the 'upper-lower' symmetry with the respect to the dot's puddle, which seems to be insensitive to gates voltage. This induces complete dephasing, or in other words, making the QD detector, even at thermal equilibrium, an extremely *accurate* path detector. In a wider aspect, the devastating effect of a common place two-state impurity nearby a coherent system may be related to this model.

We can return now to the slightly pinched QD. The main consequence of increasing the coupling to the leads of the unbiased QD is the increase of the puddle capacitance (as the puddle spreads towards the inner edge modes in the leads). This is evidenced by the smaller 'levering factor' (0.0025 vs 0.015 in the pinched dot), defined as before, $\alpha=\Delta u/\Delta V_{QDP}=C_{QDP}/C_{total}$, with $u$ the puddle's potential, $C_{QDP}$ the mutual capacitance between the plunger gate and the puddle, and $C_{total}$ the total puddle's capacitance [12]. With increasing capacitance to the inner edge mode, the added $2\pi$ per electron is shared among the outer and inner edge modes, leaving a smaller phase change in the MZI. The smaller phase fluctuations in the interferometer will result in shallower visibility dips and smaller phase lapses – as was indeed observed (Fig. 2c & 2d). Note that the increased tunneling overlap between puddle and leads (but a negligible tunneling element between the outer mode and the puddle) will modify the transmission and reflection phases of the inner edge mode, but that will have a negligible effect on the phase of the outer edge modes – hence, on the interferometer.

We return now to the non linear regime. Biasing the inner edge mode opens an energy window where the occupation of the QD is allowed to fluctuate. The conductance peaks will draw the familiar 'diamond type' pattern in the $V_{QDP}$-$V_{SD}$ plane, which 'opens up' when the levering factor is small [12]. For left-right symmetrical QD, the average occupation $P_{N+1}(V_{QDP})$ is expected to increase as the resonant level crosses $\mu_{inner}=eV_{SD}$; saturating at $P_{N+1}=0.5$ throughout the energy window; reaching $P_{N+1}=1$ when the resonant level is fully occupied. Consequently, the visibility is expected to plateau at a lower value with the MZI phase depending on the levering factor – as above (Fig. 4a). When the QD strongly deviates from left-right symmetry, transport is limited by the more opaque barrier. For a better coupled source (drain) lead, the average occupation will be $N+1$ with $P_{N+1}=1$ ($N$ with $P_N=1$), throughout the bias window. Degeneracy, $P_N=P_{N+1}=0.5$, is achieved only when the resonance level is aligned with the source (drain) electrochemical potential. In the example of Fig. 4c and 4d, with the source lead coupled better to the puddle, the visibility developed a dip and the phase lapsed when the resonant level was aligned with the source lead. The diamond-like 'visibility' is superior to that of the 'conductance', since it is provides information of the average occupation in the dot in between the energy degeneracy points, and, in particular, when the conductance of the QD is vanishingly small.

While the conductance peaks are visible only in the strong coupling regime, the visibility dips prevail throughout the entire coupling range, providing a valuable window to the otherwise inaccessible properties of the QD. Studying the conductance peaks (when visible) and the visibility dips throughout the full range of coupling strength, an interesting evolution is revealed (see Fig. 5). For relatively strong coupling of the puddle to the leads, levels' shape is Lorenzian and widths are rather broad,

$\Gamma_{elas}$~40µeV; being wider than temperature, they suggest a dwell time $\tau_{dwell}$~100pS. As the coupling of the puddle to the leads weakened, the widths narrows and saturates around $\Gamma$~12µeV - limited by the electron temperature in the leads $T=45mK$ and $\Gamma_{temp}$~3.5$k_BT$~12µeV. The dwell time was estimated from the conductance peak height; being proportional to $\Gamma_{elas}/T<1$ it was $\tau_{dwell}$~10nS. Decreasing the coupling further, the width of the visibility dip, unexpectedly, grew monotonically, reaching $\Gamma$~80µeV with no sign of saturation (Fig. 5). The rapid increase in the apparent level width, as the QD is strongly pinched (a longer dwell time), suggests an onset of decoherence in the QD, with a rate increasing as the QD was pinched off and depleted of its electrons. The level width, assumed fully inelastic, $\Gamma_{inelas}$~80µeV, reflects decoherence time of $\tau_{decoh}$~50pS when the estimated semi classical dwell time reached (by rough extrapolation) $\tau_{dwell}$>1µs. Could the decoherence result from 'back-action' of the MZI? Increasing the current in the MZI, thus increasing its shot noise, did not seem to affect the apparent width of the level - suggesting that decoherence within the QD arose from other degrees of freedom. Hence, at this regime, while the QD dephases effectively the interferometer, it ceases to be a simple quantum detector due to its own coupling to other degrees of freedom of the environment.

The assertion that decoherence is effective in a rather pinched QD was tested by observing, this time, the AB interference of a Coulomb blockaded outer edge mode. The QD was tuned to partly reflect the outer edge mode while still maintaining a measurable peak conductance. The observed visibility, $v = \xi \frac{2\sqrt{\overline{T_{QD}}}}{1+\overline{T_{QD}}}\overline{v}$, with $\overline{v}$ the visibility of the bare MZI (with open QD) and $\overline{T_{QD}}$ the effective transmission of the conductance peak, allowed to estimate the degree of coherence, $\xi$, as function of the dwell time of the outer channel in the dot (inset, Fig. 5). We observe the degree of coherence $\xi$ dropping precipitously when the dwell time increased; with nearly full decoherence at an estimated $\tau_{dwell}$~1ns. Though the dephasing processes for the inner and outer modes might be different, yet, the ubiquitous nature of self decoherence in the QD is clear [15].

An entangled system of Mach Zehnder and Fabry Perot (in a form of a quantum dot) interferometers provided a deeper look into the physics of these systems and the mutual interactions between them. A direct manifestation of the Friedel sum rule was found to be responsible, in a most striking way, to complete 'path information', and consequently, to full dephasing of the Mach Zehnder interferometer by the QD. It should be noted that electron counting in a QD is routinely now done with a coupled QPC detector [1, 3, 16]. However, the MZI, via its oscillations visibility and phase, turned out to be an extremely effective and sensitive detector of the average occupation and its fluctuations; revealing thus hidden insights of QDs. Four significant and worth noting points of the experiment should be stressed: First, each of the coupled systems, MZI and QD, served as nearly ideal 'electron detector' of the other. The detection is extremely robust and independent of many details of the experiment – thanks to fundamental physics manifested by the Friedel sum rule. Second, the QD was at equilibrium while the interferometer ran a merely infinitesimal current (only for the purpose of measuring the conductance, but not for the dephasing process). Third, while the QD fully dephased the MZI, the dual process, namely, back-action of the MZI was not evident [17]. We address this basic issue by referring to two main regimes of QD: (*i*) In the strongly pinched QD, self-decoherence and/or finite temperature dominated the broadening of the visibility dip, likely masking any back-action due to the interferometer; (*ii*) In the strongly coupled QD (with measurable conductance peaks), the back-action was expected to be weak due smaller phase lapses and shallow visibility dips in the interferometer (see Fig. 2). Fourth, the role played by the finite temperature is another point of importance. Dephasing is generally associated with energy transfer, which is possible at finite temperature. Will dephasing of the interferometer cease at 'zero' temperature, when no current flows the QD, which is tuned to unity transmission? On one hand there should not be noise in the detector (neither thermal noise nor shot noise), capable to dephase the interferometer; however, on the other hand, the interferometer may still serve as a charge detector for the QD, acting-back on the QD to induce noise, and thus leading to dephasing. We leave this issue open for further research [18].

**Acknowledgments**

We thank Oktay Göktaş for his contribution and Yuval Oreg, Yunchul Chung, Bernd Rosenow, Yoseph Imry, Izhar Neder and Assaf Carmi for helpful discussions. M. Heiblum and Y. Gefen acknowledge the partial support of the Israeli Science Foundation (ISF), the Minerva foundation, and the US-Israel Bi-National Science Foundation (BSF). M. Heiblum acknowledges also the support of the European Research Council under the European Community's Seventh Framework Program (FP7/2007-2013) / ERC Grant agreement # 227716, the German Israeli Foundation (GIF) and the German Israeli Project Cooperation (DIP).

Correspondence and requests for materials should be addressed to moty.heiblum@weizmann.ac.il.

# Figures

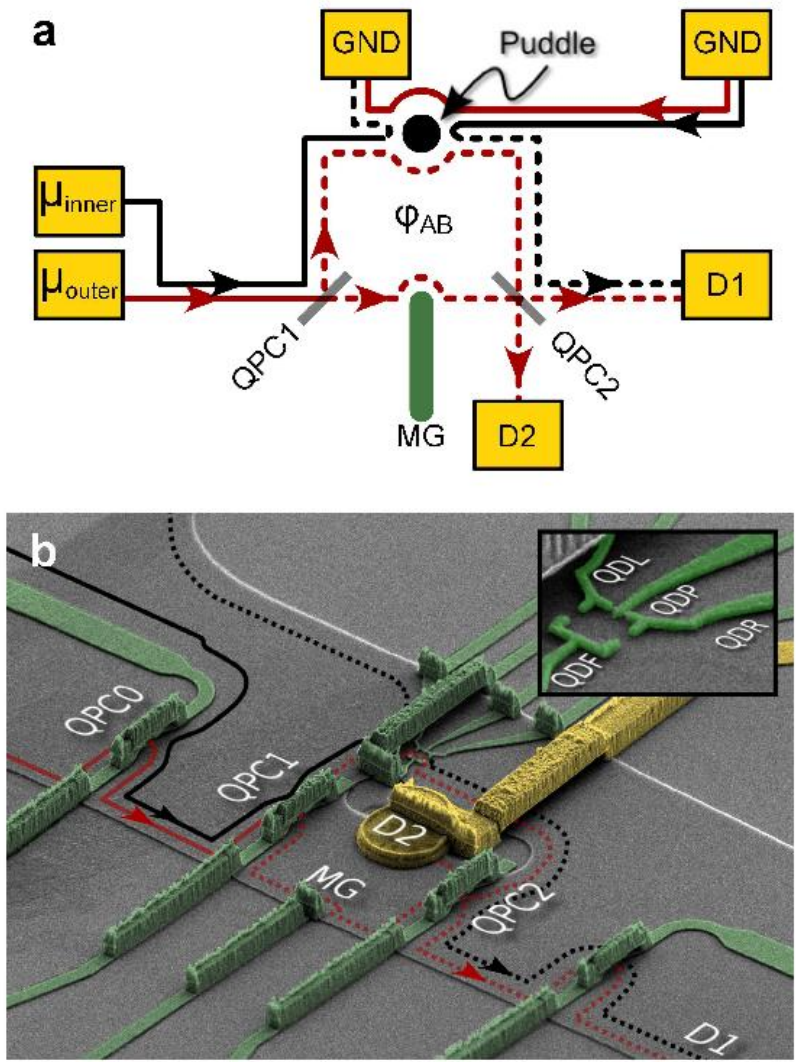

**Figure 1.** Schematics and SEM picture of the studied system. **(a)** The system comprises of an electronic Mach Zehnder interferometer (MZI) of the chiral outer edge mode in the integer quantum Hall effect (IQHE) in filling factor 2 (in red) and a quantum dot interferometer, in the Coulomb blockade regime, of the chiral inner edge mode (in black). Full lines represent full beams; dashed lines represent partitioned beams. The electron puddle lies between the two outer edge modes, one is serving as the upper arm of the MZI and the other returning from the grounded contact. The puddle is coupled capacitively and quantum mechanically to all four edge modes. When tunneling is effective, it is only via the inner edge mode. **(b)** An SEM micrograph of the fabricated structure, which was realized in a GaAs/AlGaAs heterostructure embedding a 2DEG by employing photolithography and electron beam lithography techniques. The trajectories of the outer (in red) and inner (in black) edge

modes were defined by etching and biasing surface gates (in green). Ohmic contacts serve as drain D1 (not seen) and D2 (in yellow). Measurements were conducted at an electron temperature of ~45mK and at a magnetic field of 5.6T.

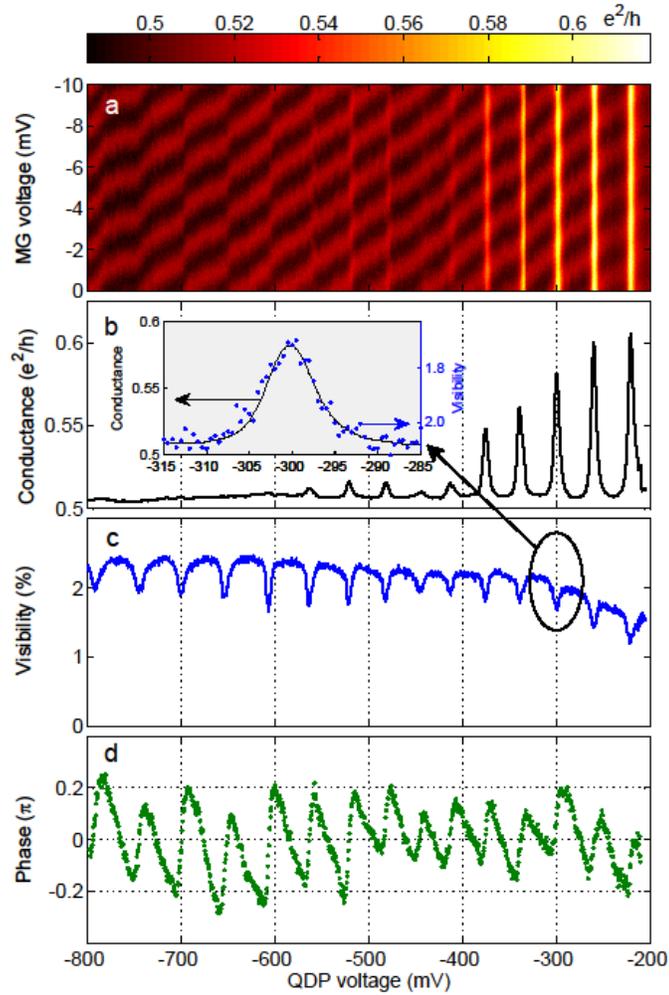

**Figure 2.** Interference pattern, visibility and phase of the MZI. **(a)** Aharonov-Bohm interference pattern of the MZI as function of modulation gate (MG) voltage and as the number of electrons in the QD (varies via the plunger gate, QDP voltage). **(b)** Coulomb blockade peaks of the QD (each peak maximum is at the degeneracy point of $N$ and $N+1$ electrons in the dot). **(c, d)** The visibility of the interference oscillations quenches and the oscillation phase undergoes a lapse at the degeneracy points. As the coupling of the QD to the leads is increased (affected inadvertently by the plunger gate voltage), the phase lapses widen and diminish while the visibility dips become shallower. Identical results were observed when the QD was at thermal equilibrium. The inset in **(b)** compares the line shape of a visibility dip with that of the corresponding Coulomb blockade peak (both Lorenzian).

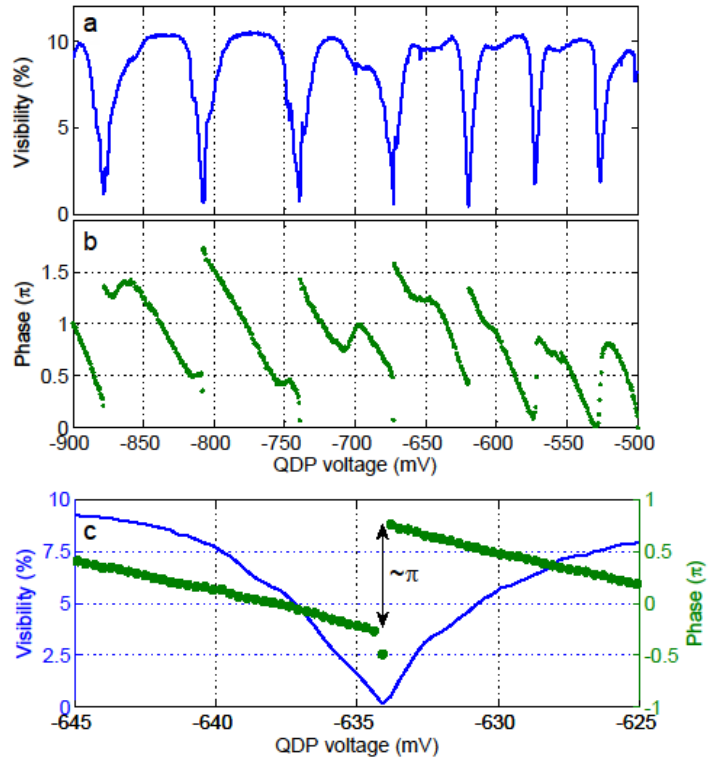

**Figure 3.** The visibility **(a)** and the oscillations phase **(b)** for a tightly pinched QD. A nearly complete loss of the visibility accompanied by abrupt lapse of π at the degeneracy points. This is a direct consequence of the Friedel sum rule [10], which ties the scattering phase of a system to its occupation. **(c)** A detailed view of the vicinity of a single degeneracy point measured at high sensitivity, exhibiting a complete visibility quench and an abrupt π phase lapse.

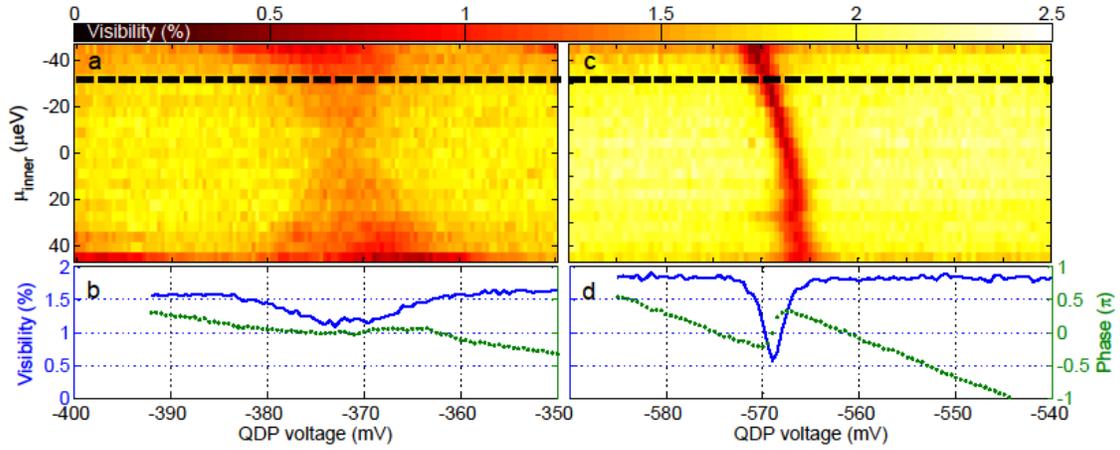

**Figure 4.** Visibility and phase in the non linear regime. Applying a DC bias to the inner edge mode opened an energy window $\mu_{inner}>\mu_{QD}>0$, for which the QD is conductive. **(a, b)** Symmetrically tuned QD ($\Gamma_{left}=\Gamma_{right}$). The visibility dips and phase evolution throughout the energy window - a cut at negative bias ~30µV exhibiting temperature smearing at both ends of the energy window. **(c, d)** Rendering the QD asymmetric, with a stronger coupling to the source, resulted in visibility quench only when the Fermi energy in the dot crossed that of the source. These results reflect directly the average occupancy QD as a function of $\mu_{QD}$: for a symmetric QD, $N+1/2$ is obtained throughout the biasing window (taking into account the level width and temperature), while for an asymmetric QD, $N+1/2$ is obtained when the Fermi energy in the dot is aligned with that of the better coupled lead.

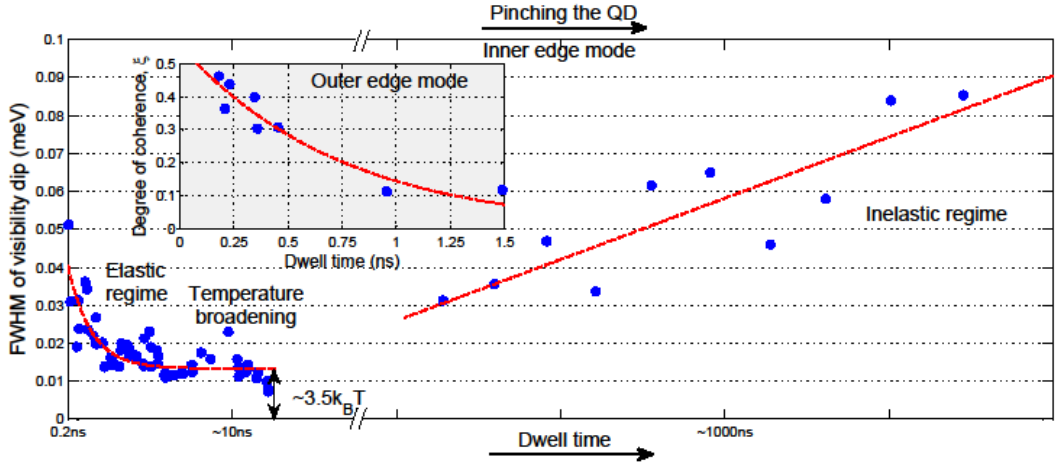

**Figure 5**. Exploring the level width of the QD in the pinched off regime. For a strongly coupled QD to the leads the width of the visibility dips follows the conductance peaks (inset, Fig 2b). The deduced typical dwell time is $\tau_{dwell}$=100ps. As the coupling weakened, with $\tau_{dwell}$~10ns, the conductance peaks and, hence, visibility dips widths became limited by the electron temperature of ~45mK. Strongly pinching off the QD, transport through the puddle in the QD diminished, leading to a long dwell time - roughly estimated $\tau_{dwell}$~1μs. The visibility dips widened at this regime, suggesting that inelastic processes come into play increasing the inelastic width, with decoherence time $\tau_{decoh}$~50ps. The inset shows the dependence of the degree of coherence of the QD on the dwell time (measured with the outer edge mode) - with the QD essentially incoherent for $\tau_{dwell}$>1ns. The red dashed lines serve as guides to the eye.